\documentclass[showpacs,10pt,twocolumn,prb]{revtex4}
\usepackage{amsmath}
\usepackage{amssymb}
\usepackage{graphics}
\usepackage{epsfig}

\begin{document}

\title{Anisotropy of Electrical Transport and Superconductivity in Metal
Chains of Nb$_{2}$Se$_{3}$}
\author{Rongwei Hu$^{1,2}$, K. Lauritch-Kullas$^{3}$, J. O'Brian$^{3}$, V.
F. Mitrovic$^{2}$ and C. Petrovic$^{1}$}
\affiliation{$^{1}$Condensed Matter Physics, Brookhaven National Laboratory, Upton New
York 11973-5000 USA\\
$^{2}$Physics Department, Brown University, Providence RI 02912\\
$^{3}$Quantum Design, 6235 Lusk Blvd., San Diego, CA 92121}
\date{\today}

\begin{abstract}
In this work we have shown bulk superconductivity and studied the anisotropy
in both the normal and superconducting states in quasi-1D conductor Nb$_{2}$%
Se$_{3}$. Electron - electron Umklapp scattering dominates electronic
transport along the direction of Nb metal chains as well as perpendicular to
it. The superconducting state is rather anisotropic with possible multi -
band features.
\end{abstract}

\maketitle

\section{Introduction}

In the recent years, there has been continuing interest in search, the
discovery and characterization of materials that exhibit exotic collective
electronic phenomena in different bonding and structure types. High T$_{C}$
cuprates, Sr$_{2}$RuO$_{4}$, some organic and heavy fermion metals are
examples of materials which exhibit low dimensional superconductivity gaps
or anisotropic Fermi surfaces.\cite{Bednorz}$^{-}$\cite{CeCoIn5} The true
character of anisotropy is one of the most important questions to be
addressed and in that context quasi-1D materials are the extreme examples in
nature.\cite{Moris}$^{-}$\cite{Sergey} Transition metal chalcogenides often
host quasi-one dimensional conducting electrons due to existence of metal
chains in their crystal structure, where band dispersion along the chain is
an order of magnitude larger than dispersion in the direction perpendicular
to chains. Intermetallic phases on the selenium-rich side of Nb-Se phase
diagram, such as NbSe$_{2}$ and NbSe$_{3}$ are fruitful model materials for
the study of low dimensional superconductivity and charge density waves
(CDW).\cite{Yokoya}$^{,}$\cite{Perucchi} On the other hand, the niobium rich
side of the Nb-Se phase diagram has been far less explored, with exception
of Nb$_{3}$Se$_{4}$ which is a superconductor with T$_{C}$ = 2.31 K.\cite%
{Ishida} In this work we show bulk superconductivity and examine the
character of anisotropy in the normal and superconducting states of Nb$_{2}$%
Se$_{3}$\cite{Igaki}, a quasi-1D conductor whose normal state electronic
transport is dominated by electron - electron Umklapp scattering.\cite%
{Rashid}

\section{Crystal Structure}

Nb$_{2}$Se$_{3}$ crystallizes in the monoclinic P$_{21/m}$ crystal structure
where Nb atoms form two types of infinite metal - metal chains running in
the b-axis direction: Nb(1) and Nb(2). Interatomic distances in the Nb(1)
chain are comparable to those in pure metal (2.97\AA ) whereas metal
distances in the Nb(2) chain are somewhat longer (3.13\AA ).\cite{Kadijk}
The origin of metal clustering in this crystal structure is due to its main
building block, M$_{2}$X$_{6}$ chains (Fig. 1) which are present in many M$%
_{2}$X$_{3}$ metal - clustering transition metal chalcogenides (M = Mo, Se,
Ta; X = S, Se).\cite{Canadell} M$_{2}$X$_{6}$ chains are formed by double
edge sharing of MX$_{4}$ chains which are in turn formed by edge sharing of
ideal MX$_{6}$ octahedra. Metal - metal bond formation across the shared
octahedral edge in MX$_{4}$ - type chains causes formation of Nb(2) chains
and distortion from ideal octahedral building blocks in Nb$_{2}$Se$_{6}$
clusters. Electrical transport properties and potential structural
instabilities in these systems are governed by the bands formed from the set
of unfilled t$_{2g}$ orbitals.

\begin{figure}[t]
%%%%%%%%%%%%%%%%%%%   F I G U R E  1  %%%%%%%%%%%%%%%%%
\centerline{\includegraphics[height=3in]{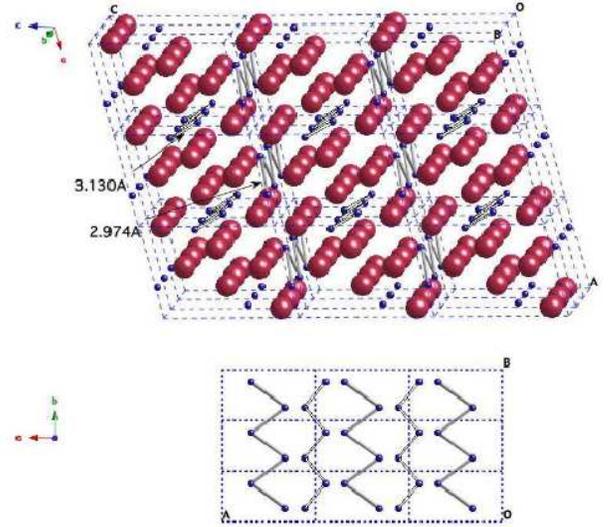}} %%%%%%%%%
%%%%%%%%%%%%%%%%%%%%%%%%%%%%%%%%%%%%%%%%%%%%%
\vspace*{-0.2cm}
\caption{Crystal structure of Nb$_{2}$Se$_{3}$ (3x3 unit cells shown) and projection onto ab plane showing two types of Nb chains, open and solid bonds. Nb(blue symbol), Sb(red symbol).}
\label{Fig1}
\end{figure}

Due to metal clustering in the basic Nb$_{2}$Se$_{6}$ building blocks band
structure is comprised of flat bands originating from Nb(2) - Nb(2) bonding
across the shared edge of the NbSe$_{4}$ chains and rather dispersive bands
formed from x$^{2}$-y$^{2}$ orbitals along the Nb(1) chain. Nb$_{2}$Se$_{3}$
is isostructural with Mo$_{2}$S$_{3}$ where metal - metal distances in both
Mo(1) and Mo(2) chains are comparable to those in pure Mo metal. Peierls
distortion associated with half filled t$_{2g}$ block bands due to the
oxidation state of Mo atoms [Mo$^{3+}$ (d$^{3}$)] doubles the unit cell so Mo%
$_{2}$S$_{3}$ exhibits CDW transition. On the other hand, there is no CDW
formation in Nb$_{2}$Se$_{3}$ since the oxidation state of Nb atoms is Nb$%
^{3+}$ (d$^{2}$). Consequently, Nb$_{2}$Se$_{3}$ is a good model material
for the study of low dimensional electronic transport along metallic chains
down to the lowest temperatures without partial destruction of the Fermi
Surface due to CDW formation, as is the case in many transition metal oxides
and chalcogenides.

\section{Experiment}

Powder X-ray patterns were taken at room temperature using a Rigaku Miniflex
with CuK$_{\alpha \text{ }}$radiation. The data were collected using 2$%
\theta $ scan in the 10$^{\circ }$ - 90$^{\circ }$ range. Several different
single crystals were oriented by a Laue camera. Electrical contacts were
made with Epotek H20E silver epoxy for current along the b-axis of the
crystal as well as perpendicular to the b axis in the a-c plane (parallel
and perpendicular to Nb - Nb chains). The electrical resistivity was
measured in a Quantum Design PPMS-9 in the temperature range from 0.4 K \ -
300K. and up to 90 kOe. The heat capacity was measured using a relaxation
technique in the same instrument. The magnetic susceptibility was measured
in a Quantum Design MPMS XL-5. The dimensions of the samples were measured
by a high precision optical microscope, the Nikon SMZ-800 with 10$\mu $m
resolution, and average values are presented. Electrical resistivity,
magnetic susceptibility and heat capacity was reproduced on several
independently grown samples from different batches in order to exclude
sample dependence.

\section{Results}

The synthesis of large single crystals allowed us to study the anisotropy in
the normal and superconducting state of Nb$_{2}$Se$_{3}$. Single crystals of
Nb$_{2}$Se$_{3}$ were grown using a molten metallic flux technique, thus
avoiding possible contamination and intercalation of transport agent atoms. 
\cite{FiskRemeika}$^{,}$\cite{CanfieldFisk}$^{,}$\cite{CanfieldFisher}
Crystals grew as thin plate-like rods with the long rod axis being the
b-axis of the crystal structure along the Nb-Nb metal chains. Crystal
structure parameters of flux grown Nb$_{2}$Se$_{3}$ crystals are in good
agreement with previously published: a=5.5051(2)\AA , b=3.4349(2)\AA ,
c=9.2369(4)\AA\ and monoclinic angle $\beta $=130.16(1)$^{\circ }$. The
anisotropy in electrical transport for current applied both along and
perpendicular to the chain direction at high temperatures is shown in Fig. 2
(a). The resistivity of our flux grown samples for current I $\uparrow
\uparrow $ b - axis ($\rho _{P}$) is in good agreement with the data from
crystals obtained using a vapor transport technique.\cite{Rashid} Electrical
transport for the current perpendicular to the chains, I $\perp $ b -axis ($%
\rho _{N}$), is up to an order of magnitude larger, implying less band
structure anisotropy than in other linear chain inorganic materials and
probably less difference in the band dispersion energy parallel to chains (b
- axis) and perpendicular to it.\cite{Oshiyama2}$^{,}$\cite{Oshiyama3} The
resistivity decreases with decrease in temperature approaching residual
resistivity values below 4K, $\rho _{P}$ = 26 $\mu \Omega $cm and $\rho _{N}$
=105 $\mu \Omega $cm. Fig. 1(b) and 1(c) show the temperature dependence of
the resistivity after subtracting the residual resistivity that has been
estimated by the extrapolation of the $\rho _{P}$ and $\rho _{N}$ curves to
T = 0.

The clustering of atoms in chains results in the quasi - one dimensional
conduction band model first proposed by Kamimura on the example of (SN)$_{x}$%
. \cite{Rashid}$^{,}$\cite{Kamimura}It was shown that electron - electron
Umklapp scattering dominates the electronic transport, whereas electron -
phonon and electron - electron normal scattering are negligible.\cite%
{Oshiyama} The temperature dependence of the resistivity is given by the
power law $\rho -\rho _{0}=CT^{n}$ (C=const.) \ For $k_{B}T\leq
(0.1-0.3)\gamma $,\ where $\gamma $ is the interchain interaction energy, $n$
takes values of $2\leq n\leq 3$, and for $k_{B}T>|\gamma |$, $n=1$. The
power law fit of both $\rho _{P}$ and $\rho _{N}$ in Fig. 2(b) is possible
below $T=15K$ whereas a linear temperature dependence is observed above $%
T=250K$. The resistivity for I $\uparrow \uparrow $ b - axis takes the form $%
\rho =\rho _{0}+CT^{n}$ with $n=2.1\pm 0.1$, \ $C=(1.40\pm \ 0.02)\times
10^{-8}\ \Omega cm/K^{n}$ and resistivity for I $\perp $ b - axis has the
same temperature dependence with $n=2.07\pm 0.04,\ C=(7.1\pm \ 0.7)\times
10^{-8}\Omega cm/K^{n}$.

\begin{figure}[t]
%%%%%%%%%%%%%%%%%%%   F I G U R E  2  %%%%%%%%%%%%%%%%%
\centerline{\includegraphics[height=3in]{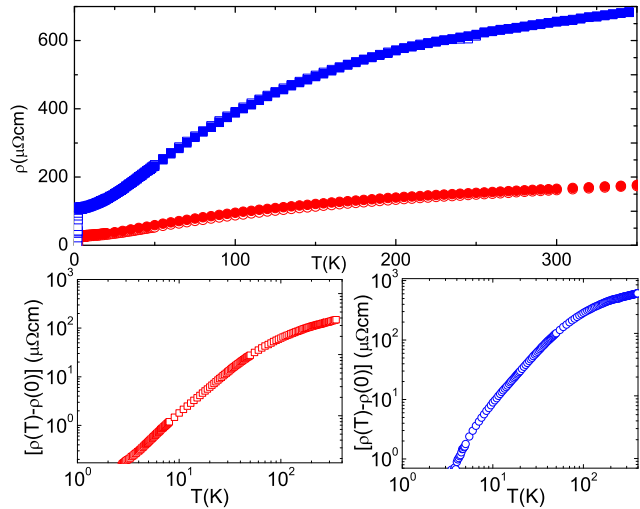}} %%%%%%%%%
%%%%%%%%%%%%%%%%%%%%%%%%%%%%%%%%%%%%%%%%%%%%%
\vspace*{-0.2cm}
\caption{(a) Electrical transport for the current applied parallel (red) and
perpendicular to Nb chains (b - axis of the crystal structure) of Nb$_{2}$Se$%
_{3}$. (b) $\protect\rho $-$\protect\rho _{0}$(T) for current applied
parallel (left) and perpendicular to Nb chains (right) }
\label{Fig2}
\end{figure}

Low temperature thermodynamic, magnetic and transport \ properties are shown
in Fig. 3. The jump $\Delta C$ at $T_{C}=2.0K$ in the specific heat and 25\%
of unsaturated -1/4$\pi $ value in M/H data at the lowest temperature of our
measurement (1.8 K) suggest a bulk superconducting transition in Nb$_{2}$Se$%
_{3}$. Electrical transport measurements show a somewhat higher transition
temperature. For current applied along the chains as well as perpendicular
to chains (b -axis) the onset of the superconducting transition is at $%
T_{\rho }^{onset}=2.4K$. At $T=2.2K$ the transition to the superconducting
state for the b - axis ($\rho _{P}$) resistivity is complete whereas
resistivity for I $\perp $ b - axis ($\rho _{N}$) shows a structure and the
change of slope. Finally, at bulk $T_{C}=2.0K$ $\rho _{N}$ is fully
superconducting. Magnetic susceptibility M/H is diamagnetic already at T$%
_{\rho }^{onset\text{ }}$and its magnitude grows towards bulk
superconductivity below T$_{C}$ = 2.0 K. By fitting the temperature
dependence of the specific heat in the normal state using $C=\gamma T+\beta
T^{3}$ we obtain $C/T=\gamma =9.96\pm 0.2mJ/moleK^{2}$ and $\theta
_{D}=223\pm 3K$ using the relation $\theta _{D}^{3}$ = $(12/5)(\pi
^{4}nR/\beta )$, where R is the gas constant and $n$ is the number of atoms
per molecule.

\begin{figure}[t]
%%%%%%%%%%%%%%%%%%%   F I G U R E  3  %%%%%%%%%%%%%%%%%
\centerline{\includegraphics[height=2.8in]{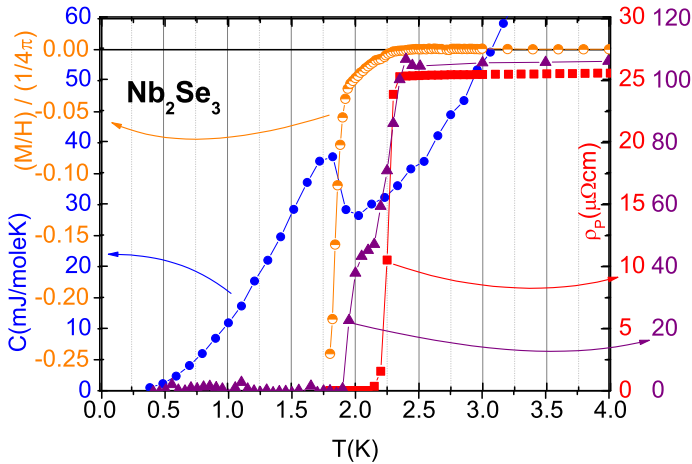}} %%%%%%%%%
%%%%%%%%%%%%%%%%%%%%%%%%%%%%%%%%%%%%%%%%%%%%%
\vspace*{-0.2cm}
\caption{Superconductivity in Nb$_{2}$Se$_{3}$. Left axis shows
thermodynamic properties (heat capacity and magnetization) and right axis
shows electrical transport properties for current applied parallel (red
symbols) and perpendicular (violet symbols) to Nb chains (b - axis of the
unit cell) }
\label{Fig3}
\end{figure}

Figure 4 presents temperature-dependent electrical resistivity data for Nb$%
_{2}$Se$_{3}$ taken at a variety of applied fields for H $\preceq $ 12 kOe
for field applied parallel and perpendicular to Nb chains and for the
current running parallel and perpendicular to Nb chains. Two features are
evident:\ there is a suppression of the superconducting phase to lower
temperatures for increasing applied field, and there is a negligible
magnetoresistivity in the normal state. The decrease of T$_{C}$ for I $%
\uparrow \uparrow $ b - axis tracks well the data for I $\perp $ b - axis
for a fixed field direction. On the other hand, suppression of
superconductivity is much stronger for H $\perp $ b axis (Nb chains) than
for a field applied along b- axis (parallel to Nb chains). A closer look at
the temperature dependent resistivity for I $\perp $ b - axis reveals a step
in the superconducting transition that persists up to 1 kOe. Using these
data, H$_{c2}$(T) curve can be deduced (Fig. 5). We notice an almost linear
temperature dependence and relatively large anisotropy of the H$_{c2}$(T)
curve for a magnetic field applied parallel and perpendicular to the Nb
chains. There is no sign of saturation down to the lowest temperature of our
measurement, 0.4 K.

\begin{figure}[t]
%%%%%%%%%%%%%%%%%%%   F I G U R E  4  %%%%%%%%%%%%%%%%%
\centerline{\includegraphics[height=3.8in]{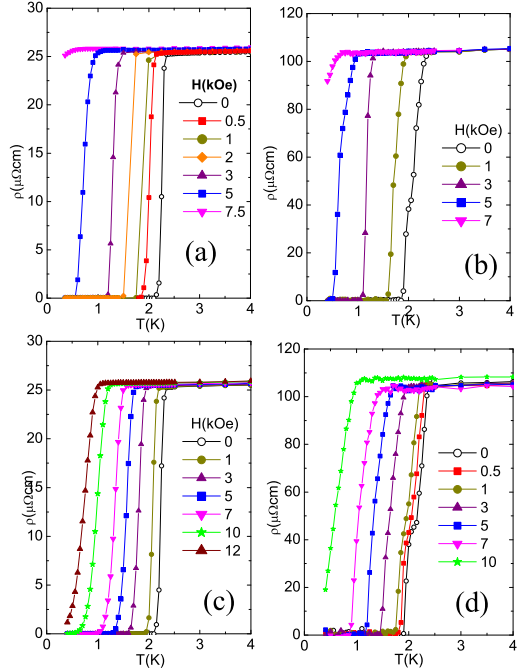}} %%%%%%%%%
%%%%%%%%%%%%%%%%%%%%%%%%%%%%%%%%%%%%%%%%%%%%%
\vspace*{-0.2cm}
\caption{Temperature dependant resistivity for field applied perpendicular
(a,b) and parallel to Nb chains (b,d), with current applied parallel (a,c)
and perpendicular (b,d) to Nb chains. }
\label{Fig4}
\end{figure}

\begin{figure}[t]
%%%%%%%%%%%%%%%%%%%   F I G U R E  5  %%%%%%%%%%%%%%%%%
\centerline{\includegraphics[height=3in]{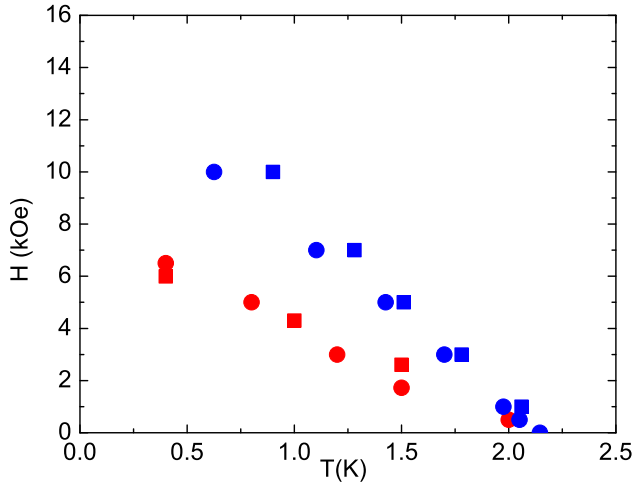}} %%%%%%%%%
%%%%%%%%%%%%%%%%%%%%%%%%%%%%%%%%%%%%%%%%%%%%%
\vspace*{-0.2cm}
\caption{H$_{c2}$(T) for fields applied perpendicular (red) and parallel to
Nb chains (blue symbols)\ with current applied perpendicular (circles) and
parallel to Nb chains (squares). }
\label{Fig5}
\end{figure}

By extrapolating H$_{c2}$(T) data to T=0 we get values of H$_{c2}$(0) = 7.5
kOe (H $\perp $ Nb chains) and H$_{c2}$(0)= 14 kOe (H $\uparrow \uparrow $
Nb chains), which is \ significantly larger. Taken as a whole, the
temperature dependence of H$_{c2}$ for Nb$_{2}$Se$_{3}$ is similar to that
found for other quasi one dimensional superconductors, for example TaSe$_{3}$%
, Nb$_{3}$Se$_{4}$ and Nb$_{3}$S$_{4}$. By using B$_{C2}$(T)=[$\Phi $0/2$\pi
\xi $(T)] we obtain coherence lengths $\xi $(0)=203\AA , $\xi $(0)=153\AA\ %
for a field applied perpendicular and parallel to Nb chains, respectively, a
bit shorter than in Nb metal $\xi $(0)=380\AA .

\section{Discussion}

Electronic transport in the normal state can be understood in the framework
of the theory of Oshiyama and Kamimura.\cite{Oshiyama} However, our results
imply that electron - electron Umklapp scattering is dominant not only along
the chain axis of a quasi-one-dimensional metal but also perpendicular to
it. The linear resistivity above 250 K and a $\rho $ $\sim $ T$^{2.1}$
temperature dependence of resistivity for both $\rho _{P}$ and $\rho _{N}$
below T = 15 K, are consistent with the possible range of the chain
interaction energy 45 K $\preceq $ $\gamma $ $\preceq $ 150 K. We note that
the power law temperature dependence of resistivity in vapor transport grown
crystals extends up to 40 K, implying a larger value of \ $\gamma $ \ than
what we obtained in our work on flux grown crystals. The difference could
arise due to the presence of other scattering mechanisms besides electron -
electron Umklapp scattering. Alternatively, this may imply that the
interchain interaction energy is sensitive to crystalline disorder and
imperfections. The flux grown crystals have a higher residual resistivity
value $\rho _{0}$ = 26 $\mu \Omega $cm for the current applied parallel to
Nb chains along b -axis of the crystal compared to crystals grown by vapor
transport reaction where $\rho _{0}$ = 0.5 $\mu \Omega $cm.\cite{Rashid}

We turn now to the properties of the superconducting state. Using McMillan's
expression\cite{McMillan}

\begin{equation}
T_{C}=\frac{\Theta _{D}}{1.45}\exp \{-\frac{1.04(1+\lambda )}{\lambda -\mu
^{\ast }(1+0.62\lambda )}\}
\end{equation}%
and value of Debye temperature from the heat capacity analysis $\theta _{D}$
= 223 K, we estimate the value of the electron - phonon coupling constant $%
\lambda $ = 1.05 assuming the empirical value of pseudopotential $\mu ^{\ast
}$ = 0.1. These results puts Nb$_{2}$Se$_{3}$ in the class of the
intermediate to strong coupling superconductors. The specific-heat jump $%
\Delta $C/($\gamma $T$_{C}$) $\approx $ 0.5 (Fig. 3) is substantially
smaller than the mean - field BCS value of 1.43.\cite{BCS} In the
weak-coupling superconductors, the reduced specific heat jump can be
interpreted in the terms of the energy gap anisotropy. The effects of an
anisotropic energy gap on the thermodynamic properties of the BCS
superconductors have been calculated in the seminal work of John Clemm:

\begin{equation}
\Delta C/(\gamma T_{C})=1.426(1-4\left\langle a^{2}\right\rangle )
\end{equation}%
where $\left\langle a^{2}\right\rangle $, the mean-squared anisotropy, is
the average over the Fermi surface of the square of the deviation of the
energy-gap parameter from its average.\cite{Clemm} Strong coupling effects
and anisotopy in general tend to work in opposite direction and the specific
heat jump would be enhanced by strong coupling. This would imply a rather
strong anisotropy of the superconducting gap in Nb$_{2}$Se$_{3}$. Equation
(2) therefore gives a rather conservative estimate of $\left\langle
a^{2}\right\rangle =0.16$. Table 1 compares the value of electron phonon
coupling parameter and the gap anisotropy for several anisotropic
superconductors.

Nb$_{2}$Se$_{3}$ exhibits zero resistivity for current applied along the b -
axis of the crystal, parallel to Nb chains ($\rho _{P}$) at T = 2.2 K,
whereas the onset of this transition \ and the transition to a diamagnetic
M/H is at T$_{\rho }^{onset\text{ }}$ = 2.4 K . However, the heat capacity
shows a transition to a bulk superconducting state at bulk T$_{C}$ = 2.0 K,
where magnetic susceptibility data for M/H shows a dive towards full Meisner
effect. These results suggest that the superconductivity of Nb$_{2}$Se$_{3}$
is a bulk effect below T$_{C}$ $\sim $ 2.0 K and filamentary in nature
between T$_{\rho }^{onset\text{ }}$and T$_{C}$. Nb$_{2}$Se$_{3}$ can be
regarded as an aggregate of one-dimensional chains that are weakly coupled
through Josephson-type junctions since the distance between the Nb atoms in
a chain is metallic but the interchain distance exceeds the metallic
distance. The temperature dependence of electrical resistivity for current I 
$\perp $ b - axis ($\rho _{N}$ - perpendicular to Nb chains) is consistent
with this interpretation. The onset temperature T$_{\rho }^{onset}$ is the
same for both $\rho _{P}$ and $\rho _{N}$, \ however $\rho _{N}$ shows zero
resistivity only at bulk T$_{C}$ and a feature indicating another
superconducting transition at T = 2.2 K that is visible in applied magnetic
field up to 1kOe. An alternative explanation for this is a small
misorientation in the current direction during electrical transport
measurement perpendicular to the Nb chains. However that would imply the
presence of two T$_{C}$'s in two different electronic substructures and
negligible interband scattering.\cite{Suhl}

$\overset{%
\begin{array}{l}
\text{Table 1: Fundamental parameters of the } \\ 
\text{superconducting state of Nb-Se and } \\ 
\text{Nb-S superconductors. Data on Nb is given } \\ 
\text{for comparison.}%
\end{array}%
}{%
\begin{tabular}{||c||c||c||c||}
\hline\hline
& T$_{C}$ (K) & $\left\langle a^{2}\right\rangle $ & $\lambda $ \\ 
\hline\hline
NbSe$_{2}$ & 7.1 & 0.04 & 0.30 \\ \hline\hline
Nb$_{3}$S$_{4}$ & 3.78 & 0.07 & 0.51 \\ \hline\hline
Nb$_{3}$Se$_{4}$ & 2.31 & 0.15 & 0.51 \\ \hline\hline
Nb$_{2}$Se$_{3}$ & 2.0 & 0.16 & 1.05 \\ \hline\hline
Nb & 9.25 & 0.01 & 1 \\ \hline\hline
\end{tabular}%
}$

\section{Conclusion}

In summary, we have showed bulk superconductivity, and have investigated the
anisotropy in electrical transport properties in the normal and in the
superconducting state of Nb$_{2}$Se$_{3}$. Our results show that Nb$_{2}$Se$%
_{3}$ is in the intermediate to large coupling limit of \ BCS theory with a
possible large mean - squared anisotropy of the energy gap on the Fermi
surface.

More microscopic measurements such as NMR, neutron scattering and tunneling
experiments would be very useful to quantify the question of possible
filamentary superconducitivity above bulk T$_{C}$ or multiband features. We
conclude that the Nb$_{2}$Se$_{3}$ is a promising model system to study
superconducting properties in quasi one dimensional metallic chain systems.

We thank S. L. Bud'ko, P. C. Canfield, M. Strongin and Z. Fisk for useful
discussions. This work was carried out at the Brookhaven National Laboratory
which is operated for the U.S. Department of Energy by Brookhaven Science
Associates (DE-Ac02-98CH10886).

\end{document}